\documentclass{aip-cp}

\usepackage[numbers]{natbib}
\usepackage{rotating}
\usepackage{graphicx}

\begin{document}

\title{Ultra-High Energy Cosmic Rays:\\ Recent Results and Future Plans of Auger}

\author[aff1]{Karl-Heinz Kampert\corref{cor1}}
\author[aff2]{the Pierre Auger Collaboration}

\affil[aff1]{Department of Physics, University of Wuppertal, Wuppertal, Germany}
\affil[aff2]{Observatorio Pierre Auger, Av. San Martin Norte 304, 5613 Malargue, Argentina;\\ full author list at: http://www.auger.org/archive/authors\_2016\_12.html
}
\corresp[cor1]{Corresponding author: kampert@uni-wuppertal.de}

\maketitle

\begin{abstract}
Over the last decade a number of important observational results have been reported using data from the Pierre Auger Observatory. We shall review some of the recent key findings that have significantly advanced our understanding of ultra-high energy cosmic rays and that have called into question the `classical' interpretation of the flux-suppression above $5\cdot 10^{19}$\,eV as being caused (solely) by the Greisen-Zatsepin-Kuzmin effect. Instead, the data suggest seeing mostly the maximum energy of extragalactic cosmic ray accelerators. This has a number of implications, ranging from reduced prospects of doing particle physics with cosmogenic neutrinos to reduced chances of seeing the sources of ultra-high energy cosmic rays at all. To address these emerging and pressing scientific questions, the Pierre Auger Observatory is presently being upgraded to AugerPrime. It will enable composition measurements into the flux-suppression region and also improve the particle physics capabilities. An Engineering Array has been deployed and is taking data. Full construction will start late 2017.
\end{abstract}

\section{INTRODUCTION}

Data from the Pierre Auger Observatory~\cite{ThePierreAuger:2015rma} and more recently the Telescope Array~\cite{AbuZayyad:2012kk} have dramatically advanced our understanding of ultra-high energy cosmic rays (UHECRs). A strong flux suppression at the highest energies, similarly to the one expected from cosmic ray energy losses in the Cosmic Microwave Background (CMB) radiation (GZK-effect)~\cite{Greisen:1966jv,Zatsepin:1966jv}, has been observed beyond any doubt~\cite{Abbasi:2007sv,Abraham:2008ru}. Moreover, strong flux limits have been placed on the photon and neutrino components at EeV energies~\cite{Abraham:2009qb,Aab:2016agp,Aab:2014bha,Abreu:2013zbq,
Aab:2015kma} disfavoring exotic particle physics models for the origin of the most energetic cosmic rays and starting to constrain the parameter space of cosmogenic neutrinos. Also, evidence is found for the presence of a large-scale anisotropy above the energy of the ankle~\cite{Abreu:2012ybu,Aab:2014ila}, and for an anisotropy on smaller angular scales at $E>5.5 \cdot 10^{19}$\,eV~\cite{Abraham:2007bb,Abreu:2010ab,PierreAuger:2014yba,Abbasi:2014lda}.

In this concert of new results, it is particularly interesting that in the data from the Pierre Auger Observatory the depth of shower maximum changes with energy in an unexpected, non-trivial way.  Around $3 \cdot 10^{18}$\,eV it shows a distinct change of $\langle X_{\rm max}\rangle$ with energy and the shower-to-shower variance decreases~\cite{Abraham:2010yv}.  Interpreted with the leading LHC-tuned shower models, this implies a gradual shift to a heavier composition \cite{Aab:2014kda,Aab:2014aea}. Recent data from the Telescope Array (TA), on the other hand, are compatible with a proton dominated composition up to $10^{19}$\,eV but also indicate a change towards a heavier composition at higher energies \cite{Abbasi:2014sfa}.

The increasing mass composition towards the highest energies, the surprisingly high level of isotropy in the arrival direction, and -- to a lesser extend -- the upper bounds of EeV photon and neutrino fluxes raise doubts about the flux suppression observed above $4 \cdot 10^{19}$\,eV as being caused (only) by the GZK-effect. Instead, a scenario in which the highest energy CR accelerators reach their limiting energy already below $E/Z \approx 10^{19}$\,eV appears to describe the bulk of data best. This interpretation would imply a radically different picture in which the maximum energy of the most powerful cosmic ray accelerators would be observed for the first time.

In the following, we shall briefly review the most recent experimental findings that lead to this new interpretation, examine their uncertainties and limitations, and discuss the results in comparison to astrophysical scenarios. The AugerPrime upgrade will address these questions by measuring the cosmic ray composition on a shower-by-shower basis into the flux suppression region, thereby enabling an exploration of the UHECR-sky individually for light and heavy primaries.

\section{THE PIERRE AUGER OBSEVATORY}

\begin{figure}[h]
\centerline{\includegraphics[width=200pt]{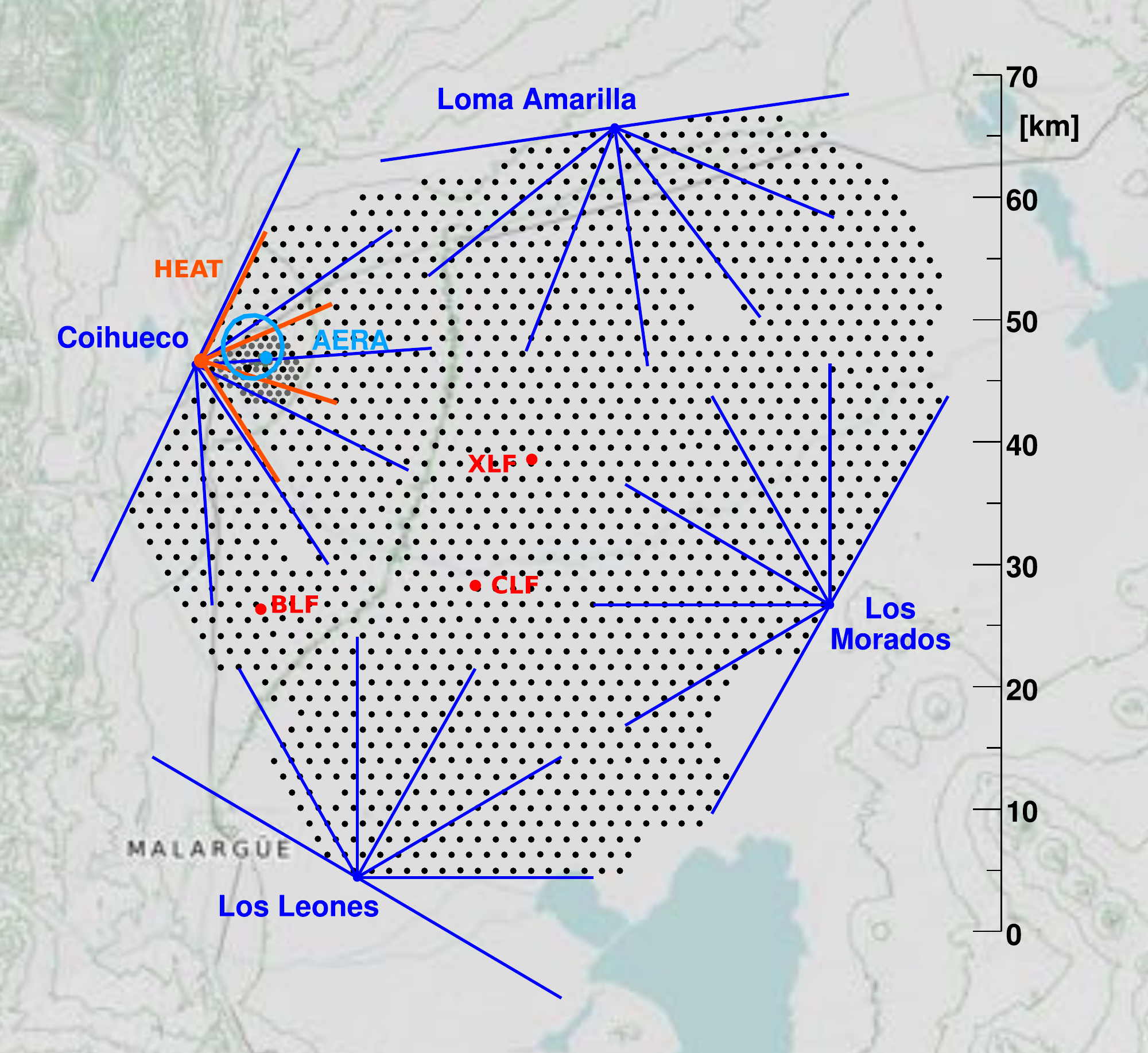}}
\caption{The Auger Observatory layout. Each dot corresponds
to one of the 1660 surface detector stations. The four fluorescence
detector sites at the periphery are shown, each with the field of
view of its six telescopes. The Coihueco site hosts three extra high elevation (HEAT) telescopes. The 750 m array is located a few kilometers
from Coihueco, as is the Auger Engineering Radio Array (AERA).\label{fig:map}}
\end{figure}

The Pierre Auger Observatory (Auger) is described in Refs.\,\cite{ThePierreAuger:2015rma,Abraham:2009pm}. It is located in Argentina (centered at $69^\circ20$ W, $35^\circ20$ S) at 1400\,m above sea level, corresponding to 870\,g\,cm$^{-2}$ of atmospheric overburden and covers an area of 3000~km$^2$ which makes it the largest cosmic ray observatory ever constructed.  
It consists of a Surface Detector array (SD) of 1600 water-Cherenkov particle detector stations (WCD) overlooked by 24 air fluorescence telescopes (FD). In addition, three high elevation fluorescence telescopes (HEAT) overlook a
surface of 23.5 km$^2$ where additional 61 WCDs (Infill-Array) are
installed (see Fig.\,\ref{fig:map}). 

Each WCD is filled with 12,000 liter of purified water. Cherenkov light from the passage of charged particles is collected by three  9'' photomultiplier tubes (PMTs) that look through windows of clear polyethylene into the water. A solar power system provides power for the PMTs and electronics package, consisting of a processor, GPS receiver, radio transceiver, and power controller. The WCD stations are placed on a triangular grid of 1500 m spacing. The 24 telescopes of the FD overlook the SD array from 4 sites: Los Leones, Los Morados, Loma Amarilla, and Coihueco (see Fig. \ref{fig:map}). Each FD site houses 6 independent telescopes, each covering a field of view of $30^\circ \times 30^\circ$ with a minimum elevation of $1.5^\circ$ above the horizon. Three additional fluorescence telescopes (HEAT) were erected at the Coihueco site. They are very similar to the original ones but can be tilted by $29^\circ$ upward to cover the elevation range from $30^\circ$ to $58^\circ$, thereby allowing a determination of the cosmic ray energy spectrum and $X_{\rm max}$ distributions in the energy range from below the second knee up to the ankle. A set of high-quality devices (Balloon Launching Facility, Central Laser Facility, Extreme Laser Facility, Lidars, etc.) is installed in the Observatory to monitor the atmospheric conditions during operation.

The infill array consists of a denser WCD array with 750\,m spacing nested within the 1500\,m array. It covers an area of 23.5\,km$^2$ and is centered 6\,km away from the Coihueco site. Full efficiency is reached from $3\cdot 10^{17}$\,eV onwards for extensive air showers (EAS) with zenith angles below $55^\circ$.

The Auger Engineering Radio Array (AERA) consists of 153 radio detection stations distributed over an area of about 17\,km$^2$. Each station is comprised of a dual polarization antenna, sensing the electric field in the north/south and east/west directions, associated analog and digital readout electronics, an autonomous power system, and a communication link to a central data acquisition system. Nine of the stations are additionally equipped with a third, vertically aligned antenna to measure the full electric field. The antennas are sensitive between 30 and 80\,MHz, chosen as the relatively radio quiet region between the shortwave and FM bands \cite{Abreu:2012pi}. 

A dedicated detector to directly measure the muon content of EAS is presently being built. The AMIGA enhancement consists of scintillator detectors buried in the soil at a depth of 280\,g\,cm$^{-2}$ next to the infill WCD stations. Each station will cover an area of 30\,m$^2$ \cite{PierreAugur:2016fvp}.

Stable data taking started in January 2004 and the baseline Observatory (without the aforementioned enhancements HEAT, infill array, AERA, and AMIGA) has been running with its full configuration since 2008 and has reached an integrated exposure of about 70,000\,km$^2$\,sr\,yr at energies above $3\cdot 10^{18}$\,eV. For each event, several observables can be reconstructed, the key ones being the energy of the primary particle, the arrival direction and, for the events detected by the fluorescence telescopes, the atmospheric depth of the air shower maximum.

\section{ENERGY SPECTRUM AND COMPOSITION}

\begin{figure}[t]
\centerline{
\includegraphics[width=210pt]{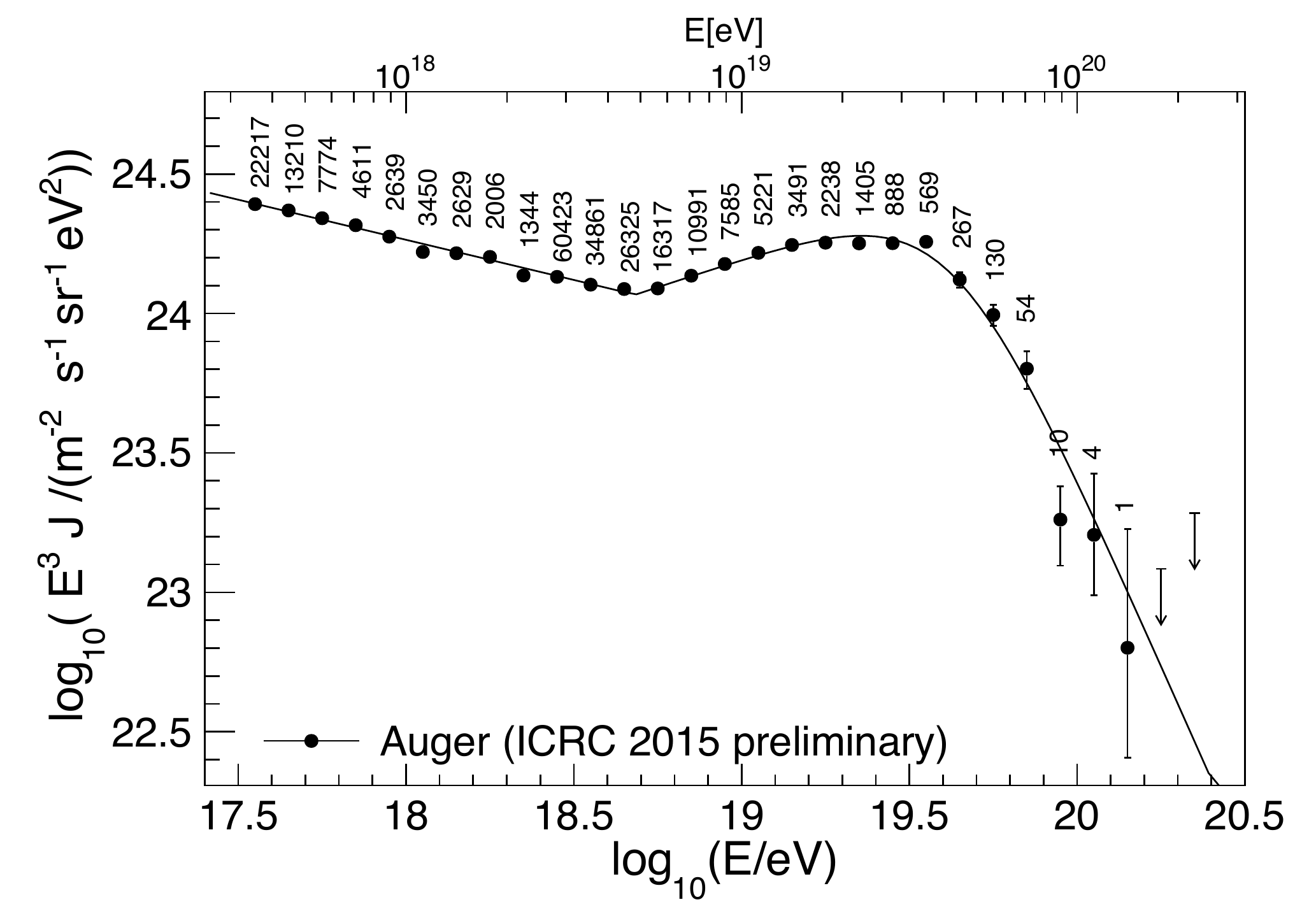}
\includegraphics[width=200pt]{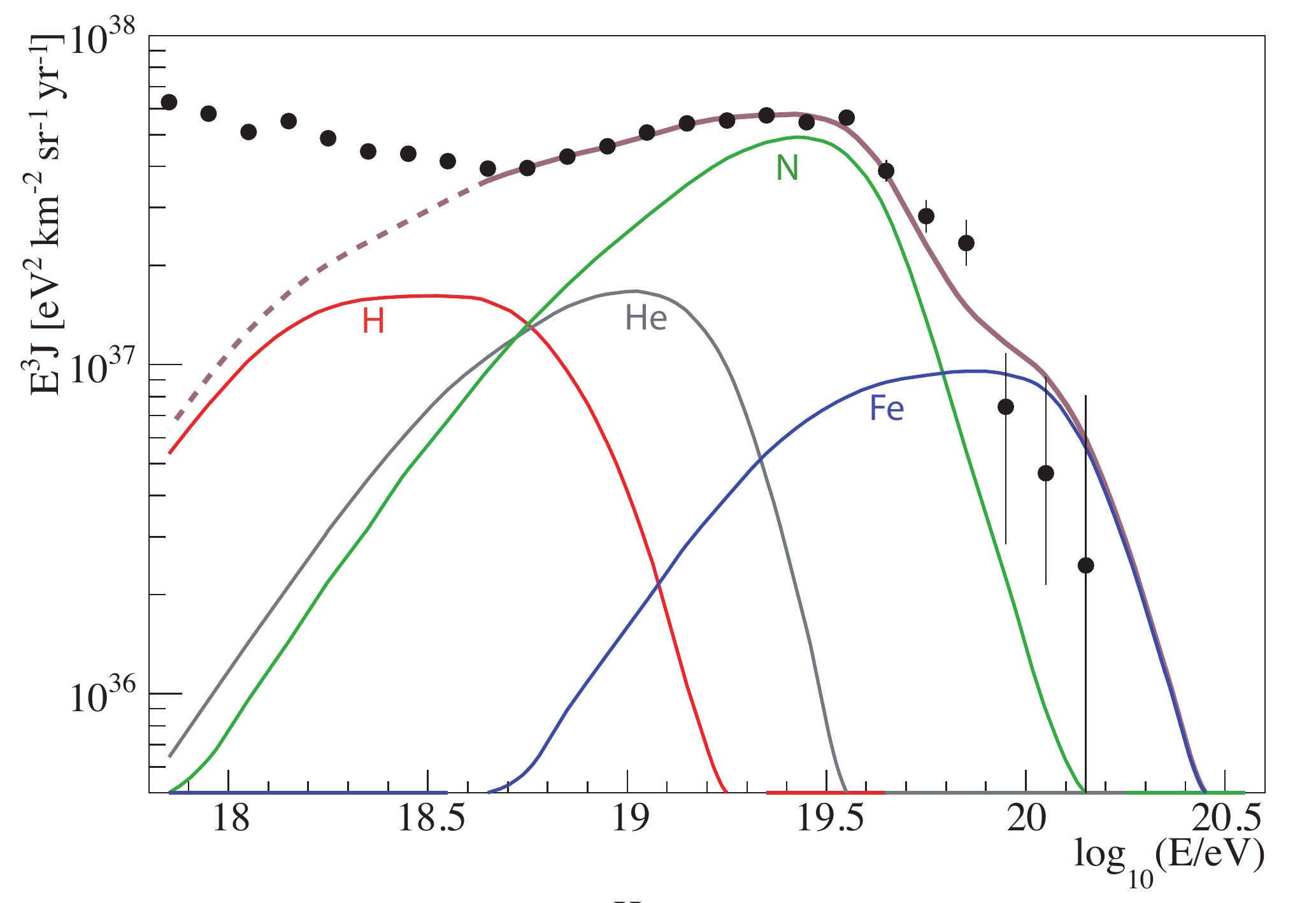}}
\caption{Left: Combined energy spectrum of cosmic-rays (multiplied by $E^3$) as measured by the Auger Observatory and fitted with a descriptive flux model. Only statistical uncertainties are shown. The systematic uncertainty on the energy scale is 14\,\%. The number of events is given above the points, which are positioned at the mean value of $\log_{10}(E/{\rm eV})$. The upper limits correspond to the 84\,\% C.L.\ (from \cite{Aab:2015bza}).
Right: Energy spectrum of UHECRs compared to the best-fit parameters for a propagation model along with Auger data points \cite{Aab:2015bza}. \label{fig:E-spectra}}
\end{figure}

The all-particle cosmic ray energy spectrum carries combined information about the UHECR sources and about the galactic and/or intergalactic media through which cosmic rays propagate. The flux suppression due to energy losses by photo-pion production and photo-disintegration in the CMB (GZK-effect) is the only firm prediction ever made concerning the shape of the UHECR spectrum. First observations of a cut-off were reported by HiRes and Auger \cite{Abbasi:2007sv,Abraham:2008ru} and in the Auger data it has reached a statistical significance of more than $20\sigma$. The most recent all-particle energy spectrum measured by the Pierre Auger Collaboration is depicted in Fig.\,\ref{fig:E-spectra}. The characteristic features have been quantified by fitting a model that describes a spectrum by a power-law below the ankle $J(E) = J_0(E/E_{\rm ankle})^{−\gamma_1}$ and a power-law with a smooth suppression at the highest energies:
\begin{equation}
J(E) = J_0 \left( \frac{E}{E_{\rm ankle}} \right)^{-\gamma_2}
\left[ 1+\left(\frac{E_{\rm ankle}}{E_s}\right)^{\Delta\gamma} \right]
\left[ 1+\left(\frac{E}{E_s}\right)^{\Delta\gamma} \right]^{-1} .
\label{eqn:flux-fit}
\end{equation}
Here, $\gamma_1$ and $\gamma_2$ are the spectral indices below and above the energy of the ankle, $E_{\rm ankle}$, respectively, $E_s$ is the energy at which the differential flux falls to one-half of the value of the power-law extrapolation from the intermediate region, $\Delta\gamma$ gives the increment of the spectral index beyond the suppression region, and $J_0$ is the normalisation of the flux, taken as the value of the flux at $E = E_{\rm ankle}$. The best fit parameters are: $E_{\rm ankle}=(4.82\pm 0.07 \pm 0.8)$\,EeV, $E_s = (42.1 \pm 1.7 \pm 7.6)$\,EeV, $\gamma_1=3.29\pm 0.02 \pm 0.05$, $\gamma_2=2.60\pm 0.02 \pm 0.1$, $\Delta\gamma=3.14\pm 0.2 \mathrm{(stat)} \pm 0.4 \mathrm{(sys)}$.

A practical definition for specifying the cutoff energy is given by $E_{1/2}$, where the flux with cutoff becomes lower by a factor of 2 than
power-law extrapolation \cite{Berezinsky:2002nc}. Computing this number as the integral of the parameterisation given by Eq.\,(\ref{eqn:flux-fit}) yields $E_{1/2} = (2.47 \pm 0.01^{\pm 0.82}_{-0.34} \mathrm{(sys)})\cdot 10^{19}$\,eV \cite{Aab:2015bza}. This value differs at the level of $3.4\sigma$ from the value of $\approx 5.3 \cdot 10^{19}$\,eV predicted in \cite{Berezinsky:2002nc} under the assumption that the sources of UHECRs are uniformly distributed over the universe and that they accelerate protons only. 

Another interpretation of the suppression region has been presented in e.g.\ \cite{Biermann:2011wf,Allard:2011aa,Taylor:2013gga,Aloisio:2013hya}. In this group of models, the flux suppression is rather than by energy losses during propagation (GZK-effect) primarily caused by the limiting acceleration energy at the sources. A good description of the Auger all-particle energy spectrum above $10^{18.6}$\,eV (to exclude contributions from galactic cosmic rays) is obtained for $E^{\rm max}_{\rm p} \simeq 6.8$\,EeV with a mix of protons and heavier nuclei being accelerated up to the same rigidity, so that their maximum energy scales like $E^{\rm max}_Z \simeq Z \times E^{\rm max}_{\rm p}$ (colored histograms in Fig.\,\ref{fig:E-spectra}, right) \cite{Aab:2015bza}. Obviously, the latter class of models (which also account for all relevant energy loss processes during propagation \cite{Kampert:2012fi,Batista:2016yrx}) leads to an increasingly heavier composition from the ankle towards the suppression region. We shall return to this aspect below. Another notable feature of such classes of models is the requirement of injection spectra being considerably harder than those expected from Fermi acceleration. This was pointed out also e.g.\ in Refs.\,\cite{Allard:2011aa,Taylor:2013gga,Aloisio:2013hya,Gaisser:2013bla}. However, as recently discussed in \cite{Mollerach:2013dza}, effects of diffusion of high energy cosmic rays in turbulent extra-galactic magnetic fields (which have not been accounted for in the aforementioned simulations) counteract the requirement of hard injection spectra ($\gamma < 2.0$) for a reasonable range of magnetic field strengths and coherence lengths. Moreover, the combined fits of energy spectrum and mass composition turn out to be quite sensitive to photodisintegration rates of nuclei in the CMB and IR-background so that better data of their photo-nuclear cross sections would be desirable.

\begin{figure}[t]
\centerline{
\includegraphics[width=400pt]{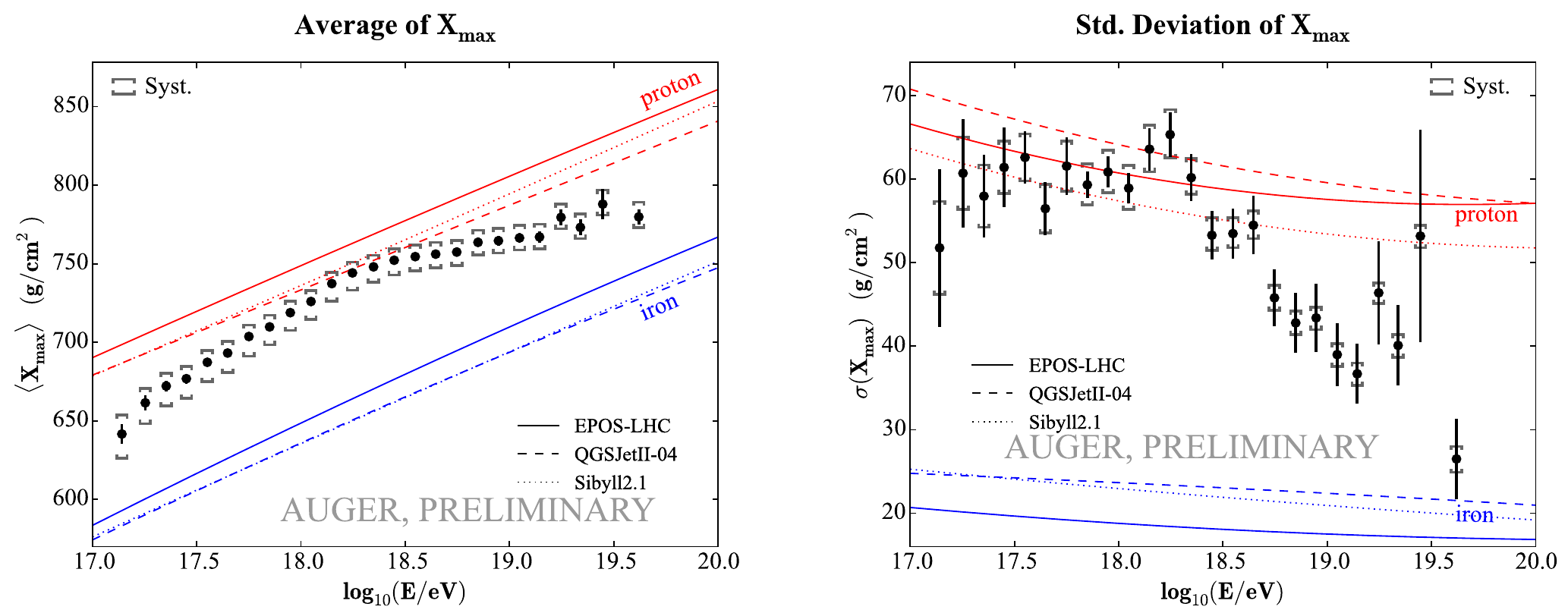}}
\caption{The mean (left) and the standard deviation (right) of the measured $X_{\rm max}$ distributions as a function of energy compared to air-shower simulations for proton and iron primaries (from \cite{Aab:2015bza}). \label{fig:Xmax}}
\end{figure}

The different interpretations of the UHECR energy spectra demonstrate the ambiguity of the all-particle energy spectrum and underline the importance of performing measurements of the primary cosmic ray composition up to the highest energies. Unfortunately, the measurement of primary masses is the most difficult task in air shower physics as it relies on comparisons of data to EAS simulations with the latter serving as a reference \cite{Kampert:2012mx,Engel:2011zzb}. EAS simulations, however, are subject to uncertainties mostly because hadronic interaction models need to be employed at energy ranges much beyond those accessible to man-made particle accelerators. Therefore, the advent of LHC data, particularly those measured in the extreme forward region of the collisions, is of great importance to cosmic ray and air shower physics and have already helped to tune the hadronic interaction models without exhibiting any bad surprises \cite{Engel:2011zzb}.

In Fig.\,\ref{fig:Xmax} (left) the mean position of the shower maximum, $X_{\rm max}$, and the width of the shower-by-shower fluctuations, $\sigma(X_{\rm max})$, are presented. $\langle X_{\rm max}\rangle$ increases between $10^{17.0}$ and $10^{18.3}$\,eV by around 85\,g\,cm$^{-2}$ per decade of energy. This value is larger than the one expected for a constant mass composition $(\sim 60$\,g\,cm$^{-2}$/decade of energy), and it indicates that the mean primary mass is getting lighter towards the ankle. Around $\sim 10^{18.5}$\,eV the observed rate of change of $\langle X_{\rm max}\rangle$ becomes significantly smaller $(\sim 26$\,g\,cm$^{-2}$/decade of energy) indicating that the composition is becoming heavier towards the flux suppression region. Similar conclusions are drawn from the fluctuations of $X_{\rm max}$ (Fig.\,\ref{fig:Xmax}, right). It should be noted that the fluctuations $\sigma(X_{\rm max})$ are sensitive also to the specific mix of primary particles. For example, a bimodal composition of equal amounts of iron and protons would show fluctuations larger than that of protons only \cite{Kampert:2012mx}. In view of this feature, it is important to note that the gradual change of primary elements suggested by the maximum-energy fit in Fig.\,\ref{fig:E-spectra} (right) also describes the observed $\langle X_{\rm max}\rangle$ and $\sigma(X_{\rm max})$ distributions of Fig.\,\ref{fig:Xmax} (see \cite{Aab:2015bza}). 

\begin{figure}[t]
\centerline{
\includegraphics[width=200pt]{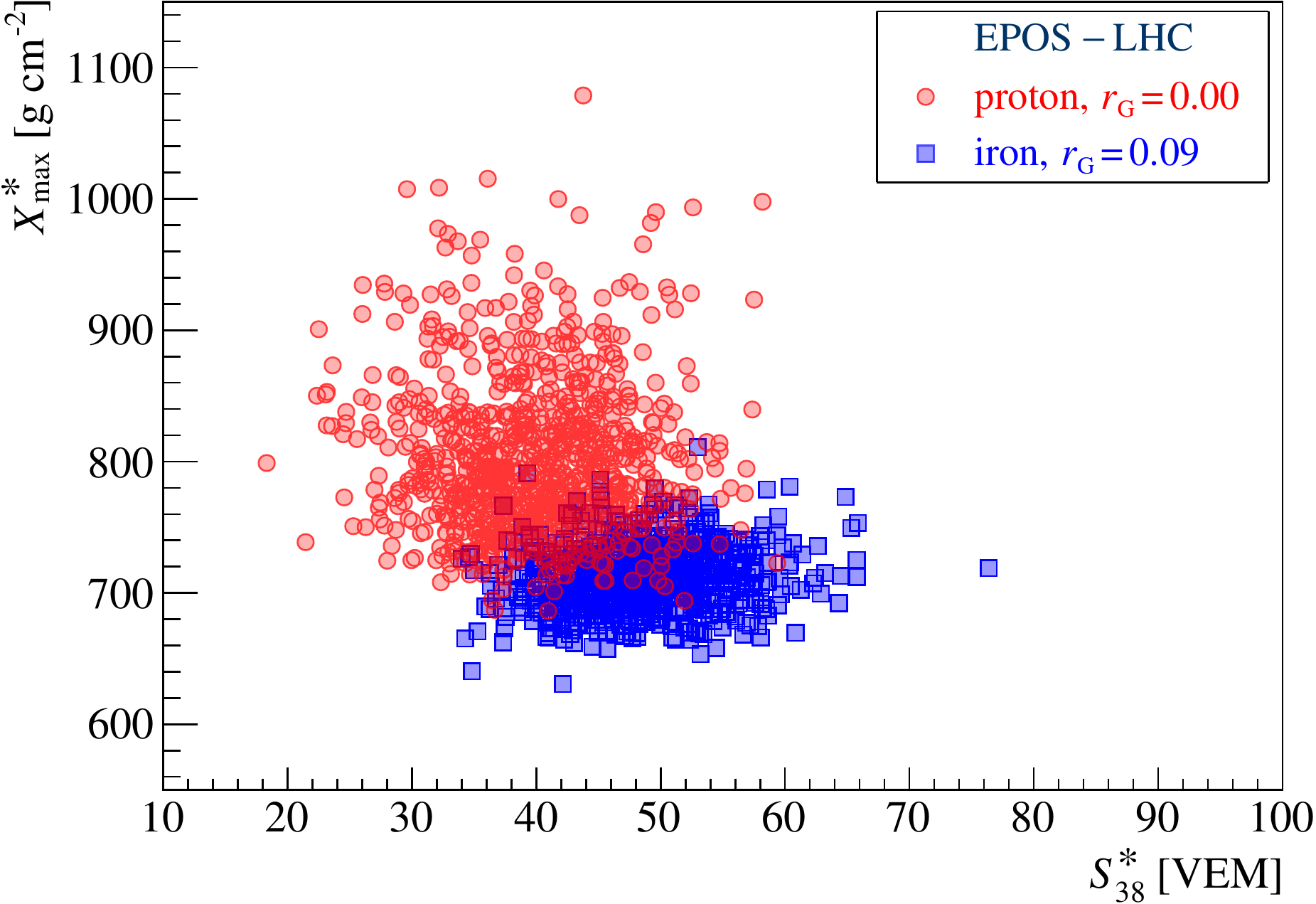}
\includegraphics[width=230pt]{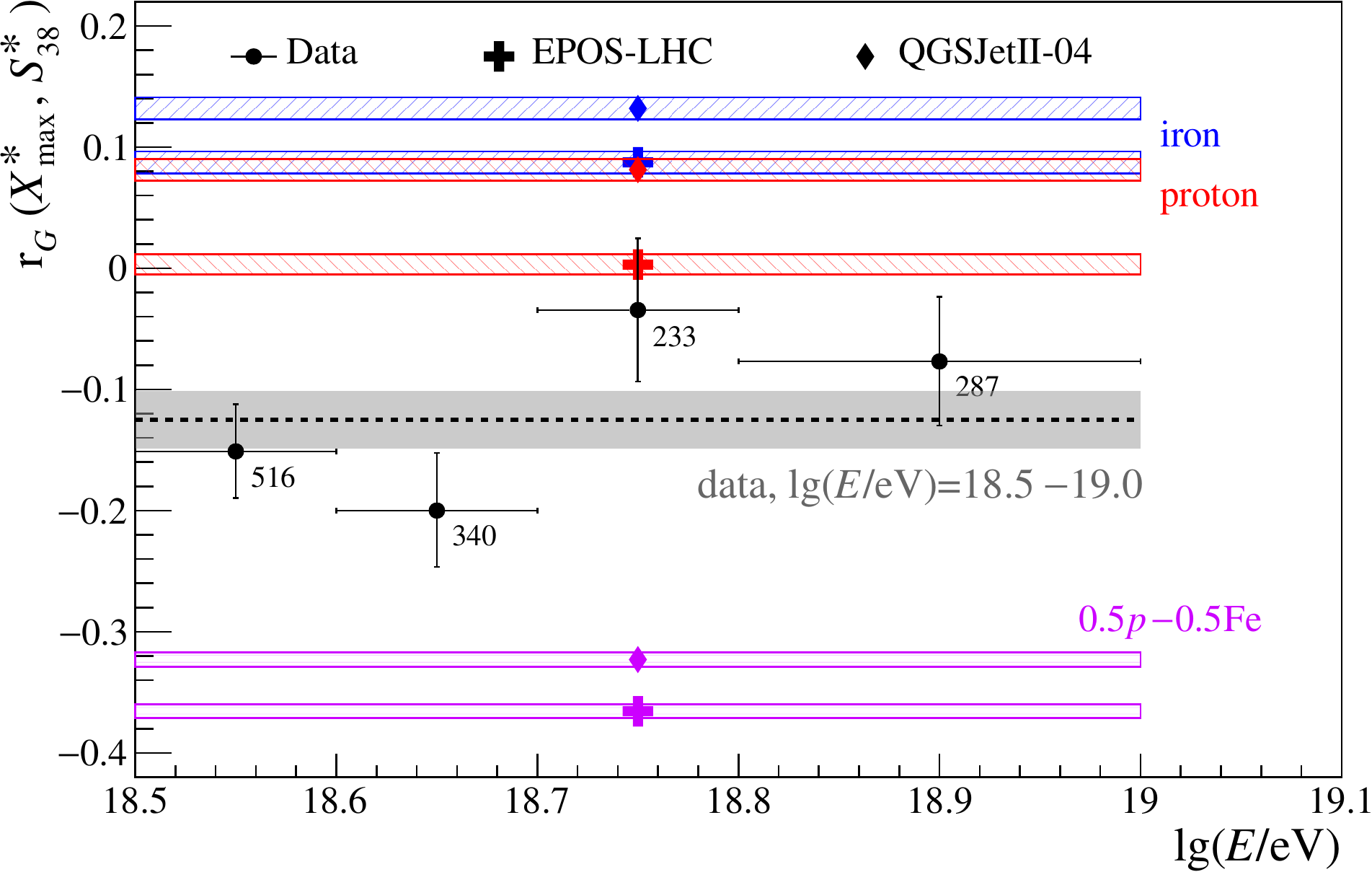}}
\caption{Left: Simulated $X_{\rm max}^*$ vs.\ $S_{38}^*$ for 1000 proton and 1000 iron showers simulated with the EPOS-LHC model of hadronic interactions. Right: Correlation coefficients $r_G$ for 4 bins of primary energies. Numbers of events in each bin are given next to the data points. The gray band shows the measured value for data in the whole range lg($E$/eV) = 18.5 to 19.0. Predictions for the correlations $r_G$ in this range for pure proton and iron compositions, and for the extreme mix 50\,\%\,$p$ plus 50\,\% Fe from EPOS-LHC and QGSJetII-04 models are shown as hatched bands. The widths of the bands correspond to statistical errors (from \cite{Aab:2016htd}). \label{fig:mixed-ankle}}
\end{figure}

The composition measurements by the fluorescence telescopes suggest a light composition at the ankle and a gradual change towards higher energies. Such a departure from a proton-dominated composition has been unexpected based on results of first generation UHECR observatories. Moreover, the assumption of a proton dominated composition has motivated the so-called ``proton dip-model'' \cite{Berezinsky:2002nc}. In this model, both the flux suppression at the highest energies and the shape of the ankle are described by energy loss processes of protons in the CMB, namely by photo-pion and electron-positron production in the CMB, respectively. This prediction requires an almost pure proton composition at the ankle. A sensitive and robust tool to measure the purity of the primary mass composition has been presented very recently in \cite{Aab:2016htd}. It employs the correlation between the depth of shower maximum, $X_{\rm max}$, and the signal in the water-Cherenkov stations, $S(1000)$, of EAS registered simultaneously by the fluorescence and the surface detectors of the Pierre Auger Observatory. Such a correlation measurement is a unique feature of a hybrid air-shower observatory with sensitivity to both the electromagnetic and muonic components and allows an accurate determination of the spread of primary masses in the cosmic-ray flux. 

As can be seen in Fig.\,\ref{fig:mixed-ankle} (left), a pure Fe composition exhibits a positive correlation between $X_{\rm max}^*$ and $S_{38}^*$ and a pure proton composition a correlation near to zero. Since $S(1000)$ and $X_{\rm max}$ of an air shower depend on the energy and, in the case of $S(1000)$, also on its zenith angle, we scale $S(1000)$ and $X_{\rm max}$ to a reference energy and zenith angle. In this way, a decorrelation between the observables from combining different energies and zenith angles in the data set is avoided. We scale $S(1000)$ to $38^\circ$ and 10\,EeV using the calibration curves from \cite[A Schulz]{ThePierreAuger:2013eja} and $X_{\rm max}$ to 10\,EeV using an elongation rate of 58\,g\,cm$^{-2}$/decade. These scaled quantities are marked with an asterisk: $X_{\rm max}^*$ and $S_{38}^*$.
As a measure of the correlation between $X_{\rm max}^*$ and $S_{38}^*$ the ranking coefficient $r_G(X_{\rm max}^*$, $S_{38}^*)$ is taken. In Fig.\,\ref{fig:mixed-ankle} (right) the observed values of $r_G$ are presented in four individual energy bins. The data are consistent with a constant negative value $r_G = -0.125 \pm 0.024$ (stat). Simulations for any pure composition with EPOS-LHC, QGSJetII-04, and Sibyll 2.1 models give $r_G \ge 0.00$ and are in conflict with the data. Equally, simulations for all proton-helium mixtures yield $r_G \ge 0.00$. However, the observations are naturally explained by a mixed composition including nuclei heavier than helium $A > 4$, with a spread of masses $\sigma(\ln A) \simeq 1.35 \pm 0.35$. This leads us to the conclusion that the mass composition at the ankle is not pure but instead mixed so that proposals of almost pure compositions, such as the aforementioned dip-scenario, are disfavored as the sole explanation of UHECR. Along with the Auger results discussed above, our findings indicate that various nuclei, including masses $A > 4$, are accelerated to ultrahigh energies $(> 10^{18.5}$\,eV) and are able to escape the source environment.

\section{SEARCHES FOR NEUTRAL PARTICLES AND MAGNETIC MONOPOLES}

Searches for neutral particles, such as ultra-high energy (UHE) photons, neutrinos, and neutrons are amongst the natural methods used to unravel the mystery of the origin of cosmic rays of the highest energy. Protons and nuclei interacting with the CMB during propagation and suffering photo-pion and electron-pair production energy loss processes are expected to produce a flux of UHE photons and neutrinos by the decay of neutral and charged pions, and bremsstrahlung, respectively. Photons at $E>1$\,EeV can propagate for a few tens of Mpc without being absorbed while neutrinos can travel to the observer with no interaction or deflection. The expected cosmogenic fluxes depend on the composition and maximum energy of CRs at the sources and the emissivity, distribution, and cosmological evolution of the acceleration sites. Thus, observing UHE photons or neutrinos, can pose constraints on the UHECR origin and properties of the sources.

\begin{figure}[t]
\centerline{
\includegraphics[width=245pt]{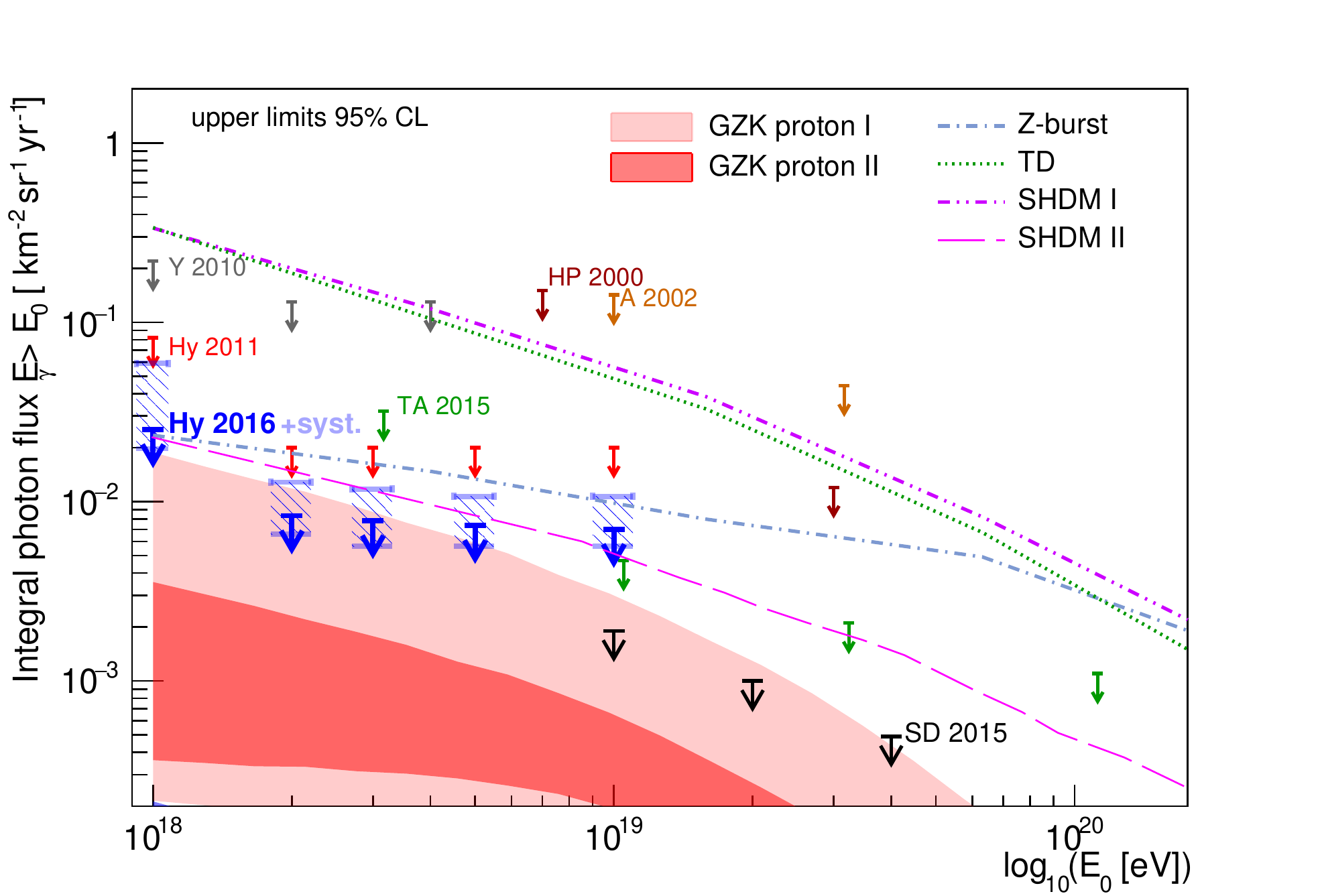}
\includegraphics[width=230pt]{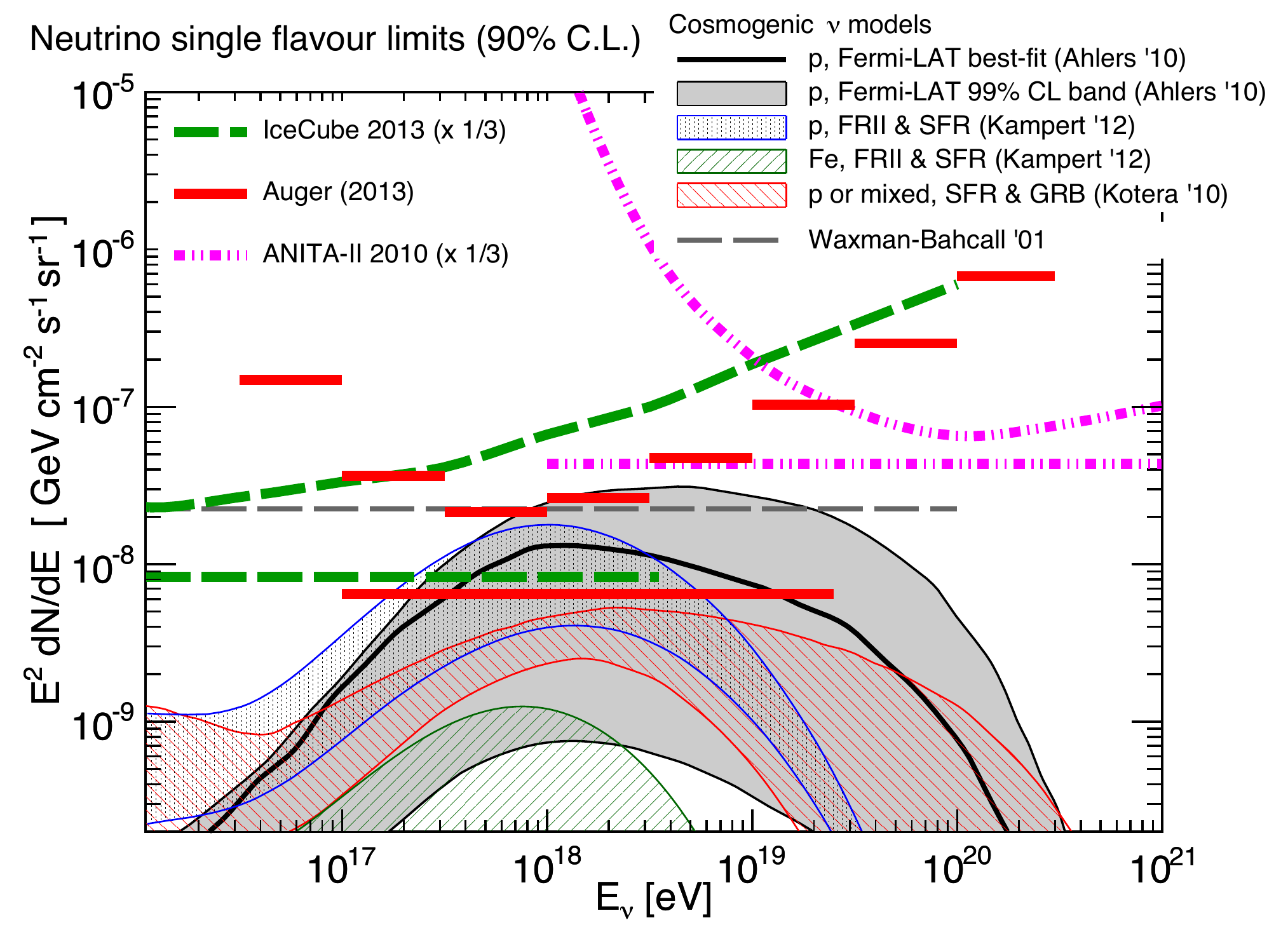}}
\caption{Left: Upper limits at 95\,\% C.L.\ to the diffuse flux of UHE photons derived from Auger SD and hybrid data (for References see \cite{Aab:2016agp}). Right: Upper limits to the diffuse flux of UHE neutrinos at 90\,\% C.L.\ in integrated (horizontal lines) and differential form. Limits from Auger (red lines) are compared with cosmogenic neutrino models (for References see \cite{Aab:2015bza}). All neutrino limits and fluxes are converted to a single-flavour. \label{fig:photons-neutrinos}}
\end{figure}

{\bf UHE Photons:} Showers induced by photons are characterised by a lower content of muons and larger average depth of maximum of longitudinal development ($X_{\rm max}$) than showers initiated by nuclei with the same energy. This is due to the radiation length being more than two orders of magnitude smaller than the mean free path for photo-nuclear interaction, causing a reduced transfer of energy to the hadron/muon channel, and to the development of the EAS being delayed by the typically small multiplicity of electromagnetic interactions. The Landau-Pomeranchuk-Migdal (LPM) effect becomes important beyond 10 EeV \cite{Mcbreen:1981yc} and is accounted for in the CORSIKA \cite{Heck98a} simulations. The CORSIKA output is then injected to the Auger detector simulation package Offline \cite{Abreu:2011fb} to study the detector response to photon induced EAS. Searches for photons at $E>10$\,EeV are performed both by hybrid data and by data from the SD only. The latter data set has more statistical power but less discrimination per event as compared to the hybrid data. For this reason, the upper limits derived from the hybrid data-set reach down to lower photon energies while the limits derived from the SD data set dominate at the higher energies. Recent updates of the diffuse flux limits of photons derived from Auger data are presented in Fig.\,\ref{fig:photons-neutrinos} (left) together with expected diffuse photon fluxes originating from the GZK-process or from particle physics motivated top-down models \cite{Aab:2016agp}. These photon limits are the most stringent ones currently available above 1\,EeV and they disfavour top-down models most strongly and also start to constrain the parameter space for cosmogenic photons in case of pure proton sources \cite{Kampert:2012mx}.

{\bf UHE Neutrinos:} The complementary search for neutrinos exploits their extremely small cross-section with matter. At large zenith angles $(\theta > 60^\circ)$ the thickness of the atmosphere traversed is large enough to absorb almost completely the electromagnetic component of EAS initiated by nucleons or even photons, leaving their signal dominated by muons. EAS initiated by neutrinos very deep in the atmosphere, on the other hand, have a considerable amount of the electromagnetic component remaining (``young'' showers). Two types of neutrino-induced showers are considered:
(1)\,Earth-Skimming (ES) showers ($90^\circ < \theta < 95^\circ$, induced by $\nu_\tau$ travelling upward with respect to the vertical at the ground) can skim the crust of the Earth and interact close to the surface, producing a $\tau$-lepton which can decay in flight in the atmosphere close to the SD. At $10^{18}$\,eV the mean decay length of the $\tau$-lepton is $\approx 50$\,km. (2)\,Downward-Going (DG) showers ($60^\circ < \theta < 90^\circ$) initiated by neutrinos of all flavours interacting in the atmosphere close to the SD through neutral current or charged current interactions, as well as showers produced by $\nu_\tau$ interacting in the mountains surrounding the observatory.

To identify neutrinos we search for very inclined ``young'' showers. Inclined showers are identified by: (i)\,a large ratio length/width (L/W) of the major/minor axis of the ellipse encompassing the footprint of the EAS and (ii)\,the distribution of apparent speeds of the trigger time between stations being required to have an average value close to the speed of light and a small RMS. Large values of the Area-over-Peak (AoP) ratio in the time traces of the WCD indicate a large contribution of the electromagnetic component. For all the channels the observable used to identify neutrinos is generally based on the AoP of stations. The full selection strategy is described in \cite{Aab:2015kma}. The region for neutrino candidates is defined using a training data sample. Based on the distribution of the data in the training set, the cut in the separation variable is set such, that only one background event is expected in 50 yrs of data taking. After unblinding, no candidate event has been observed in the data sample which yields the upper limits depicted in Fig.\,\ref{fig:photons-neutrinos} (right). The Auger Observatory is the first air shower array to set a limit below the Waxman-Bahcall flux \cite{Bahcall:1999yr} and it starts to constrain the fluxes predicted under a range of assumptions for the composition of the primary flux, source evolution, and model for the transition from galactic to extragalactic cosmic-rays. For example, and as shown in Fig.\,\ref{fig:photons-neutrinos} (right), cosmogenic $\nu$-models that assume a pure primary proton composition and strong cosmological source evolution (FRII-type) and that are still constrained by the GeV observations of Fermi-LAT are disfavoured by our data. 

\vspace*{1mm}{\bf UHE Neutrons:} Data from the Auger Observatory have also been used to search for high energy neutrons from galactic sources. Searches for neutron point sources are motivated by the fact that any proton source is expected to produce neutrons by charge-exchange interactions in which a $\pi^+$ takes the positive charge of the proton and a leading neutron emerges with most of the energy that the proton had. A flux of neutrons from a single direction can then be detected as an excess of air showers arriving from that direction within the angular resolution of the Observatory. The mean decay path length for a neutron is $9.2\,(E$/EeV)\,kpc. Thus, above 1\,EeV, the Galactic center is within the mean decay length, and above 2\,EeV most of the Galactic disk is within range for neutron astronomy. A $1/E^2$ differential energy spectrum of protons from TeV to EeV in some of the H.E.S.S. sources would produce a neutron flux that is readily detectable by Auger. However, as reported in \cite{Aab:2014caa}, searches so far do not find evidence for a neutron flux from any class of galactic candidate sources. A positive detection of neutron sources would have identified sources of EeV protons in the Galaxy. 

\vspace*{1mm}{\bf Ultrarelativistic Monopoles:} Grand Unified Theory (GUT) models predict the production of supermassive magnetic monopoles ($M \approx 10^{26}$\,eV/$c^2$) in the early Universe at the phase transition corresponding to the spontaneous symmetry breaking of the unified fundamental interactions. When the original unified group undergoes secondary symmetry breaking at lower energy scales, so-called Intermediate-Mass Monopoles (IMMs, $M \sim 10^{11} - 10^{20}$\,eV/$c^2$) may be generated. These particles, too massive to be produced at accelerators, may be present today as a cosmic radiation relic of such early phase transitions. An ultra-relativistic IMM would deposit a large amount of energy in its passage through the Earth's atmosphere, comparable to that of an UHECR. For example, an IMM of $E=10^{25}$\,eV with $\gamma = 10^{11}$ loses $\approx 700$\, PeV/(g\,cm$^{-2}$) which sums up to $\approx 10^{20.8}$\,eV of deposited energy when integrated over an atmospheric depth of $\approx 1000$\,g\,cm$^{-2}$. When compared with a standard UHECR proton shower of energy $10^{20}$\,eV, the IMM shower presents a much larger energy deposit and deeper development, due to the superposition of many showers uniformly produced by the IMM along its path in the atmosphere. This distinctive feature has successfully been used in the analysis of the FD and SD data and has provided the most stringent upper limits so far for $\gamma > 10^9$ \cite{Aab:2016poe}. This result is valid for a broad class of intermediate-mass ultrarelativistic monopoles ($E_{\rm mon} \approx 10^{25}$\,eV and $M \sim 10^{11}-10^{16}$\,eV/c$^2$) which may be present today as a relic of phase transitions in the early Universe.

\section{ANISOTROPIES}

Finally, we shall discuss anisotropies in the arrival direction of UHECR, because important information about the nature and origin of UHECR is contained in the distribution of arrival directions. Measurements of the arrival directions of cosmic ray events are practically free from systematic errors. However, for searches of large scale anisotropies detector stability is of key importance. Apart from the (unknown) distribution of sources over the sky, the two main factors that determine the UHECR anisotropy are deflections in cosmic magnetic fields and attenuation due to the interactions with the background radiations.

The extragalactic magnetic fields are known quite poorly and are inferred mostly from Faraday rotation measurements of the light from extragalactic sources. In typical $B$-fields, a proton of $10^{20}$~eV would be deflected by $\lesssim 2^\circ$ over a distance of $50$~Mpc which is supported by many simulations. The Galactic magnetic field is known much better. This field would deflect a proton of $10^{20}$~eV by about $2-4^\circ$ depending on the direction. The deflections in the random component of the Galactic field were argued to be subdominant \cite{Tinyakov:2004pw,Pshirkov:2013wka}.

As is clear from the above numbers, if primary particles are predominantly protons, one might expect to recover the distribution of sources over the sky, with possibly bright spots of the size of a few degrees corresponding to individual bright sources. This is particularly true in presence of the GZK-effect which limits the horizon to a few ten Mpc. On the other hand, if primary particles are intermediate or heavy nuclei, the flux distribution should be anisotropic in a manner similar (but not identical) to the source distribution at the scale of a few tens of degrees, but all of the small-scale structure would be washed out. Note that because of the small propagation distance, at the highest energies the sources are expected to be distributed anisotropically due to the large-scale structure of the Universe \cite{Kampert:2014eja}.

\begin{figure}[t]
\centerline{
\includegraphics[width=220pt]{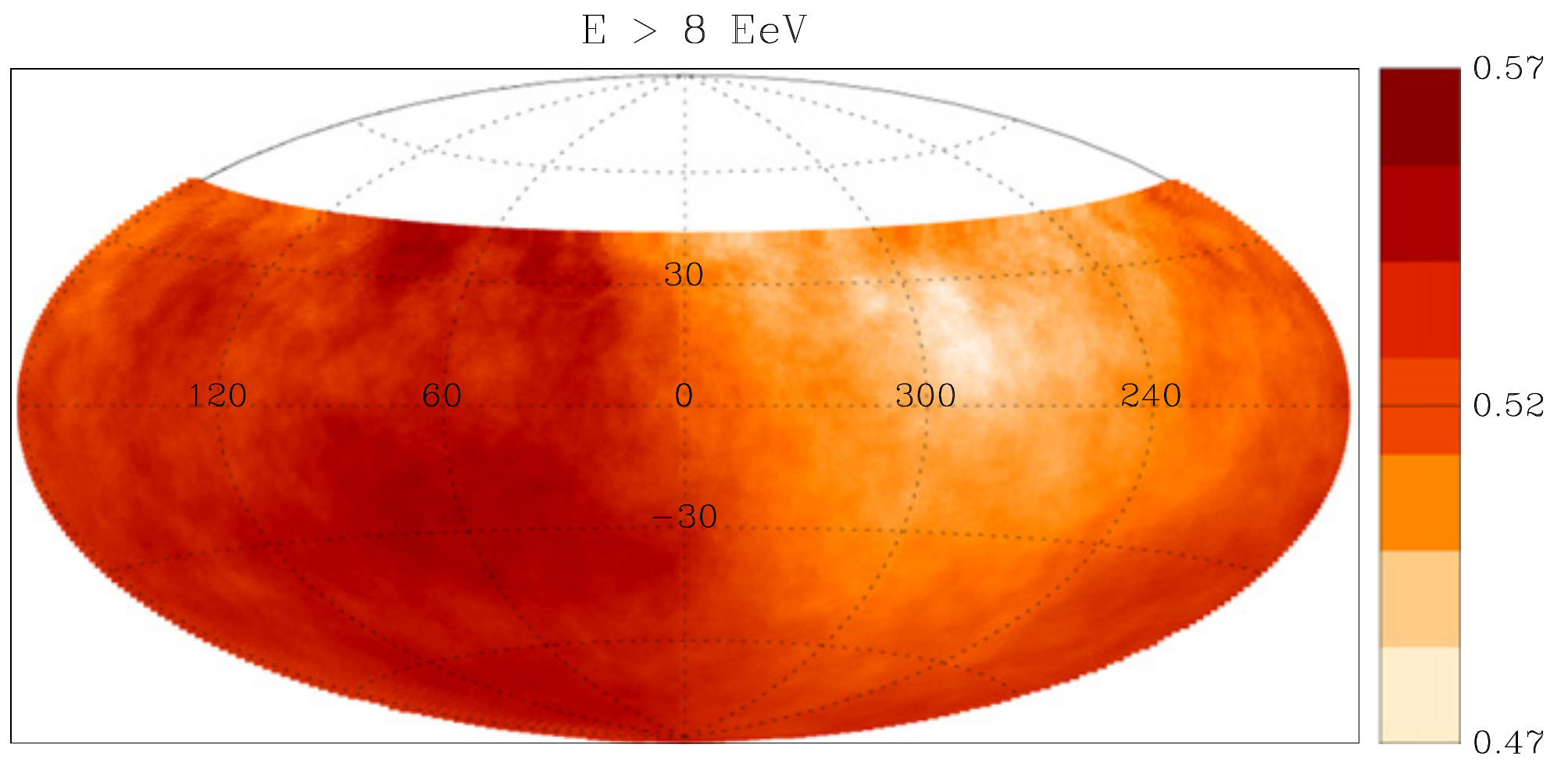}
\includegraphics[width=220pt]{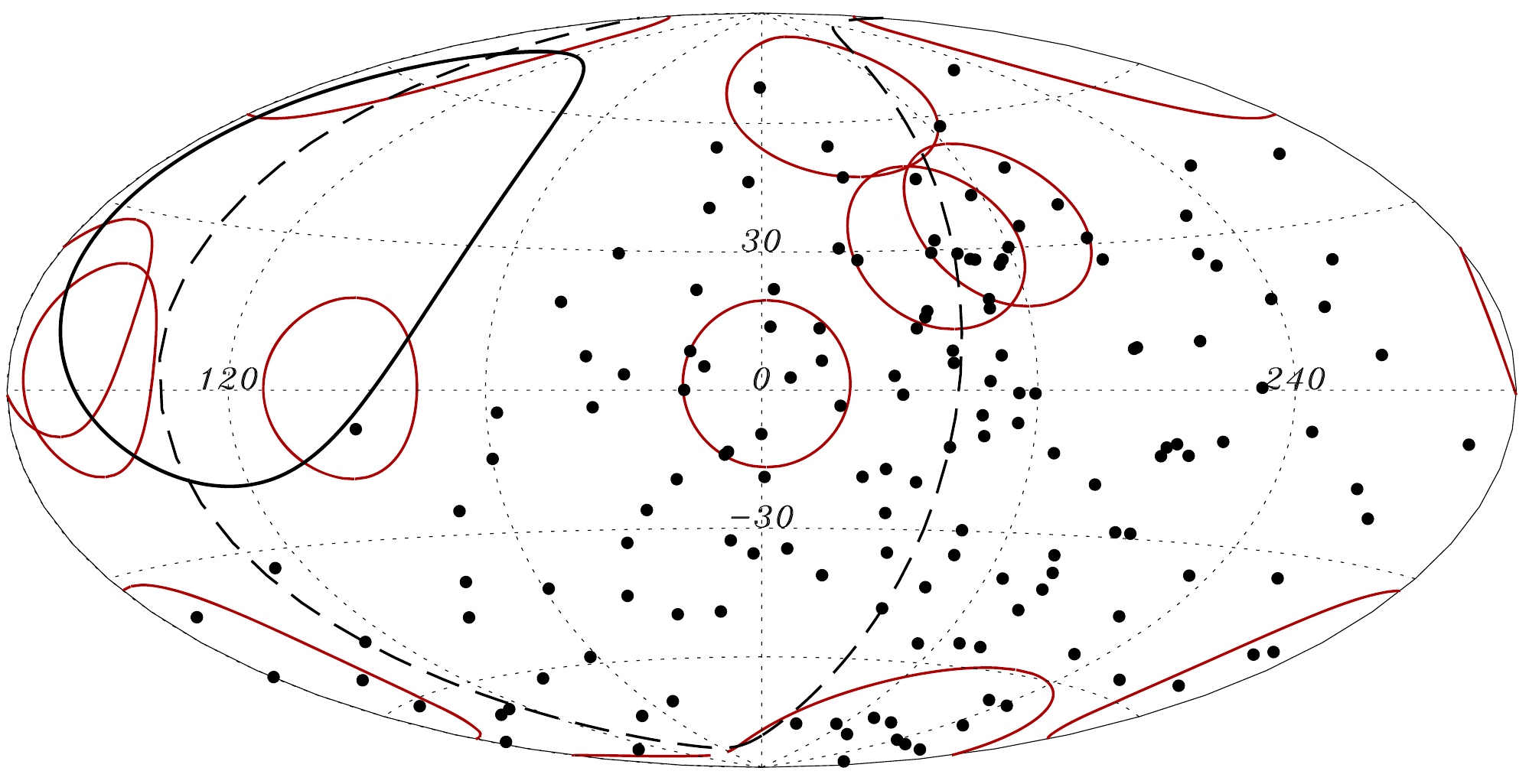}}
\caption{Left: Sky map in (equatorial coordinates) of flux, in km$^{-2}$\,yr$^{-1}$\,sr$^{-1}$ units, smoothed in angular windows of $45^\circ$ radius, for observed events with energies $E >8$\, EeV (from \cite{ThePierreAuger:2014nja}). Right: The sky map (in Galactic coordinates) shows the events with $E \ge 58$\,EeV, together with the Swift AGNs brighter than $10^{44}$\,erg\,s$^{-1}$ and closer than 130\,Mpc, indicated by red circles of $18^\circ$ radius (from \cite{PierreAuger:2014yba}). \label{fig:anisotropy}}
\end{figure}

The Pierre Auger Collaboration (also jointly with the TA Collaboration \cite{Aab:2014ila}) has performed a number of anisotropy searches on different angular scales by applying different techniques.
Indications for deviations from isotropic expectations were found but for the time being no results exceeded a significance of $5\sigma$. 

Large-scale angular modulations of the flux have been studied by performing a Rayleigh harmonic analysis on the right ascension distribution. In the energy bin between 4 and 8\,EeV, the harmonic coefficients are consistent with zero within their uncertainties with no evidence for departures from isotropy in the right ascension distribution. However, in the highest energy bin, $E>8$\,EeV, used already in previous analyses \cite{Abreu:2012ybu}, the first harmonic has an amplitude $r_1 = (4.4 \pm 1.0) \cdot 10^{-2}$, yielding a chance probability $P(\ge r_1) = 6.4\cdot 10^{-5}$ \cite{ThePierreAuger:2014nja}. The amplitude of the second harmonic is less significant, with a 2\,\% probability of arising by chance.  The observed dipole amplitude suggests a large-scale anisotropy is imprinted on the CR arrival directions of extragalactic CRs towards $\simeq 95^\circ$ in right ascension and could result from the diffusive propagation of extragalactic cosmic rays in the extragalactic turbulent magnetic field. This could happen if the amplitude of the field is large and/or if the cosmic rays have a component with large electric charge \cite{Harari:2013pea}. A large-scale anisotropy is also expected in the case of small magnetic deflections if the cosmic ray sources are distributed similarly to the matter in our local neighbourhood.

At higher energies (around and above the cutoff in the spectrum) the situation is more complex. The Auger collaboration has reported some excess of events with $E>58$~EeV around the direction towards the Centaurus supercluster at a distance of about 60~Mpc and Centaurus A, the closest radio-loud active galaxy at a distance of about 4~Mpc. The largest excess was found for a circular region of $15^\circ$ in which 14 events are observed while 4.5 are expected from isotropy \cite{Aab:2015bza}. In the Northern sky, the TA collaboration has also observed a deviation from isotropy in the data set with $E>57$~EeV at similar angular scales in the direction about $20^\circ$ from the Supergalactic plane, with no evident astrophysical structures in the closer vicinity \cite{Abbasi:2014lda}. The significance of this so called ``hot spot'' has been estimated at the $3.6\sigma$-level. It is intriguing to see the same energy threshold and a similar angular scale for both local overdensities. The angular scale appears larger than expected for protons from nearby sources. However, it is too early to draw firm conclusions from these interesting observations and more data need to be collected. The Auger Collaboration has also performed cross-correlation analyses with bright AGN from catalogs. For Swift-BAT bright AGNs, the correlation maximises for $D = 130$\,Mpc and $\mathcal{L} > 10^{44} $\,erg\,s$^{-1}$, with a threshold energy of $E_{\rm th} = 58$\,EeV and an angular radius $\Psi = 18^\circ$. For those parameters, 62 pairs are observed between 155 cosmic rays and 10 AGNs (with $\mathcal{L}_X > \mathcal{L}_{\rm min}$) while 32.8 are expected from isotropy. A sky map is shown in Figure\,\ref{fig:anisotropy} (right) representing these events and AGNs in galactic coordinates. The penalized probability to find in isotropic simulations stronger correlations under the same scan on ($\Psi$, $E_{\rm th}$, $\mathcal{L}_{\rm min}$, $D$) is $P \simeq 1.3$\,\% \cite{PierreAuger:2014yba}.

\section{UPGRADE TO AUGERPRIME}

The results obtained so far, some of which are presented in the previous sections, have dramatically advanced our understanding of UHECR. However, it is still not possible to determine whether the observed flux suppression at the highest energies is due to the GZK-effect or due to a limiting acceleration power of the sources. It is evident that this puzzle must be resolved in order to identify sources or source regions. The key lies in better identification of the primary composition, especially extending to the highest energies. An event-by-event understanding of the identities of the particles will elevate the quality of several analyses, including the anisotropy study by comparing sky-distributions of  low- and high-$Z$ particles, or by seeking evidence of a maximum acceleration energy ($E \propto Z$) by observing the primary mass increasing as the flux declines. Moreover, explicit experimental confirmation of the existence of even a small ($\sim 10$\,\%) flux contribution of light elements at the highest energies, as reported in \cite{Aab:2014aea}, will be a decisive ingredient for assessing the physics potential of existing and future cosmic ray, neutrino, and gamma-ray observatories.

\begin{figure}[t]
\centerline{
\includegraphics[width=220pt]{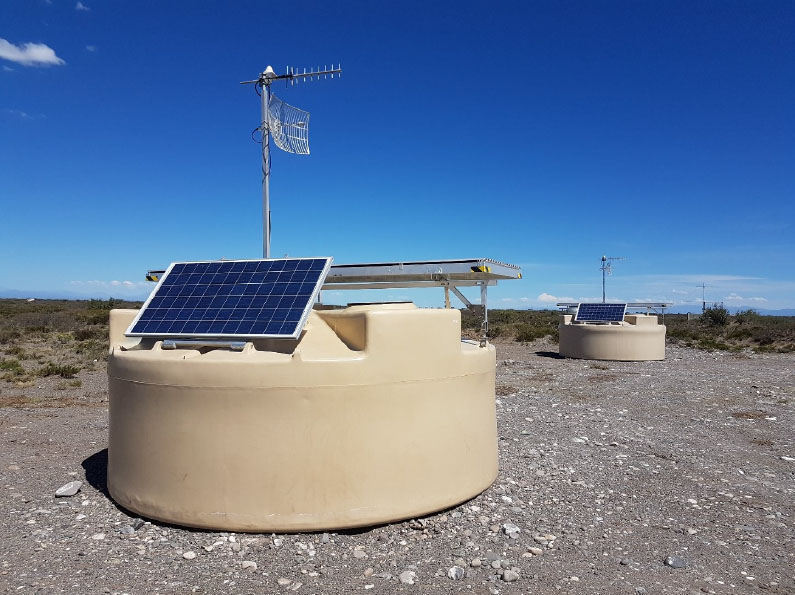}
}
\caption{Two AugerPrime detector stations installed in the Pierre Auger Observatory and arranged in a twin configuration for test purposes. The SSD is mounted on top of the WCD. \label{fig:AugerPrime}}
\end{figure}

The most promising way to obtain further composition-sensitive information is the discrimination between the electromagnetic and muonic components of the EAS with ground-array measurements. As a result of intense R\&D-efforts the Collaboration has decided to complement each WCD by 3.8\,m$^2$ large scintillators (SSD: Scintillator Surface Detector) to be placed on top of the tanks. This will enhance the electron/muon-discrimination of the SD and aid the reconstruction of the primary masses on shower-by-shower basis based on their electron-to-muon ratio. A direct measurement of the muon content will be obtained for a sample of showers with 61 muon detectors deployed on a 750 m grid in the infill area of the SD. Each of these scintillation detectors will cover an area of about 30\,m$^2$ and will be buried about 2 meters underground.
AMIGA will be used to verify and fine-tune the methods used to extract the muon component of showers with the SSD. Moreover, a significant increase of the duty cycle of the fluorescence telescopes is planned by operating the FD also when a large fraction of the moon illuminated. During
such operations the PMT gain will be reduced by lowering the HV to avoid high anode current and, therefore, an irreversible deterioration of the
PMT sensitivity. Details of the AugerPrime upgrade are described in \cite{Aab:2016vlz}. An Engineering Array of 12 stations has been installed in September 2016 and is since successfully taking data. Its purpose is to verify the performance characteristics, the stability over an extended period, and the integration into the communication and data-streams of the Pierre Auger Observatory. All design specifications are met and partly even exceeded so that full construction will start late 2017 and finish within 2 years after that.

\section{Conclusions and Outlook}

The Pierre Auger Observatory continues to provide a wealth of new data of unprecedented quality which already boosted the understanding of UHE cosmic rays. The features in the UHECR energy spectrum -- the ankle and the suppression at the highest energies -- have been established beyond any doubt. The spectral slopes before and after the ankle have been measured to the second digit. However, the parameters of the break at the highest energies are known less accurately. The position of the break is compatible with the GZK cutoff for protons, but other explanations, most prominently a scenario where we observe the limiting energy of the sources, are favored by the spectral shape. The strongest support for this interpretation stems from the increasingly heavy composition at energies above the ankle. Moreover, the sky is surprisingly isotropic even at the highest energies and only first indications for anisotropies at angular scales of about $15-20^\circ$ are found both by TA and Auger. These results are difficult to understand for pure proton compositions unless extremely strong galactic and extragalactic magnetics fields are assumed.

The interpretation of the flux-suppression region has profound implications also for the understanding of the ankle. Seeing the GZK-effect would naturally allow to explain the ankle in terms of $e^+e^-$-pair production losses in the CMB while the maximum energy scenario relates the ankle to the transition of galactic to extragalactic cosmic rays. The novel analysis presented in Fig.\,\ref{fig:mixed-ankle} (right) favours a mixed composition at the ankle, such as is expected from the maximum-energy scenario and rules out a pure composition at the ankle which is required by the $e^+e^-$ pair production explanation. This concert of results has important consequences also for predicting cosmogenic neutrino and photon fluxes and thereby also for the design criteria for future experiments. 

Addressing these problems by collecting just more UHECR data would be inefficient and would hardly solve the issues. For this reason, the Pierre Auger Collaboration is presently upgrading its Observatory to extend the measurements of primary masses on an event-by-event basis up to the highest energies. This enables the reconstruction of mass-selected sky distributions. Selecting light primaries at the highest energies may in fact be the only way to detect point sources, thereby opening the door to cosmic ray astronomy. Simulations demonstrate that a possible subdominant proton contribution at the highest energies pointing back to their sources within a few degrees would be detected much faster, i.e.\ with less total exposure, as in cases where the protons are hidden in the almost isotropic sky of the heavier primaries \cite{Aab:2016vlz}. 

Moreover, improving the separation of the electromagnetic and muonic shower component will also enhance the detection capabilities for high energy photons and neutrinos and enable stringent tests of hadronic interaction models at energies much higher than available at the LHC \cite{Collaboration:2012wt,Aab:2016hkv}.

\section{ACKNOWLEDGMENTS}
Its a pleasure to thank the organizers of the Carpathian Summer School of Physics 2016 for inviting me to participate in this stimulating meeting. The successful installation, commissioning, and operation of the Pierre Auger Observatory would not have been possible without the strong commitment and effort from the technical and administrative staff in Malarg\"ue. Financial support of KHK by the German Ministry of Research and Education (Grant 05A14PX1) and by the Helmholtz Alliance for Astroparticle Physics (HAP) is gratefully acknowledged.


\nocite{*}
\bibliographystyle{aipnum-cp}%
\bibliography{khk-references}

\end{document}